\documentclass[%
 aip,
 amsmath,amssymb, nofootinbib,
 reprint,%
]{revtex4-1}

\usepackage{graphicx}
\usepackage{dcolumn}
\usepackage{bm}

\usepackage[utf8]{inputenc}
\usepackage[T1]{fontenc}
\usepackage{mathptmx}

\usepackage{bm}
\usepackage{braket}
\usepackage{mathtools}

\usepackage[usenames,dvipsnames]{color}

\begin{document}

\title{Benchmarking Semiclassical and Perturbative Methods for Real-time Simulations of Cavity-Bound Emission and Interference}

\author{Norah M. Hoffmann}
\thanks{These two authors contributed equally}
\email{norah-magdalena.hoffmann@mpsd.mpg.de}
\affiliation{Department of Physics, Max Planck Institute for the Structure and Dynamics of Matter and Center for Free-Electron Laser Science, 22761 Hamburg, Germany}
\author{Christian Sch\"afer}
\thanks{These two authors contributed equally}
\email{christian.schaefer@mpsd.mpg.de}
\affiliation{Department of Physics, Max Planck Institute for the Structure and Dynamics of Matter and Center for Free-Electron Laser Science, 22761 Hamburg, Germany}
\author{Niko S\"akkinen}
\email{nsakkinen@gmail.com}
\affiliation{Department of Physics, Max Planck Institute for the Structure and Dynamics of Matter and Center for Free-Electron Laser Science, 22761 Hamburg, Germany}
\author{Angel Rubio}
\email{angel.rubio@mpsd.mpg.de}
\affiliation{Department of Physics, Max Planck Institute for the Structure and Dynamics of Matter and Center for Free-Electron Laser Science, 22761 Hamburg, Germany}
\author{Heiko Appel}
\email{heiko.appel@mpsd.mpg.de}
\affiliation{Department of Physics, Max Planck Institute for the Structure and Dynamics of Matter and Center for Free-Electron Laser Science, 22761 Hamburg, Germany}
\author{Aaron Kelly}
\email{aaron.kelly@dal.ca}
\affiliation{Department of Physics, Max Planck Institute for the Structure and Dynamics of Matter and Center for Free-Electron Laser Science, 22761 Hamburg, Germany}
\affiliation{Department of Chemistry, Dalhousie University, Halifax, Canada, B3H 4R2}

\date{\today}

\begin{abstract}
We benchmark a selection of semiclassical and perturbative dynamics techniques by investigating the correlated evolution of a cavity-bound atomic system to assess their applicability to study problems involving strong light-matter interactions in quantum cavities. The model system of interest features spontaneous emission, interference, and strong coupling behaviour, and necessitates the consideration of vacuum fluctuations and correlated light-matter dynamics. We compare a selection of approximate dynamics approaches including fewest switches surface hopping, multi-trajectory Ehrenfest dynamics, linearized semiclasical dynamics, and partially linearized semiclassical dynamics. Furthermore, investigating self-consistent perturbative methods, we apply the Bogoliubov-Born-Green-Kirkwood-Yvon hierarchy in the second Born approximation. With the exception of fewest switches surface hopping, all methods provide a reasonable level of accuracy for the correlated light-matter dynamics, with most methods lacking the capacity to fully capture interference effects.\\
\end{abstract}

\date{\today}

\maketitle

\section{Introduction}

Profound changes in the properties of cavity-bound molecular systems can be achieved in regimes where the quantum nature of light becomes important. A few notable examples are the change of conductivity in semiconductors due to vacuum field hybridization \cite{Orgiu2015}, the appearance of mixed states due to strong coupling \cite{Chikkaraddy2016,Casey2016}, and multiple Rabi splittings caused by ultrastrong vibrational coupling \cite{George2016}. \textcolor{black}{Although the forefront of the rapidly expanding domain of cavity-modified chemistry has been strongly driven by experiments, theoretical investigations have offered complementary insights into the various possibilities opening up with this new field of research\cite{AddR1, AddR2, Add3, Add4, Add5,Add6,Feist2015,Schachenmayer2015,Cirio2016,Flick2017,ruggenthaler2018quantum,schafer2019modification}.} 

To describe chemical processes that are strongly correlated with quantum light\cite{thomas2016,hiura2018cavity,thomas2019tilting}, requires an accurate and flexible, furthermore computationally efficient, treatment of the light-matter interactions. Thus, in order to meet the demand of developing an \textit{ab initio} theoretical description of cavity modified chemical systems, extensions to the traditional theoretical tool-kits for quantum optics and quantum chemistry are required. Therefore, in this paper we focus on semiclassical dynamics methods, which due to the simplicity, efficiency, and especially scalability, present an interesting alternative or extension to existing quantum electrodynamical wavefunction\cite{galego2015,Flick2017a,luk2017multiscale,schafer2018ab} and density-functional (QEDFT) based approaches\cite{ruggenthaler2014,Pellegrini2015,flick2017ab,ruggenthaler2018quantum}. 

The semiclassical concept has the advantage of providing an intuitive qualitative understanding of the dynamics through trajectories in phase space. Furthermore, many semiclassical methods do not exhibit an exponential scaling of the computational effort with system size or simulation time. However, these methods can fail to quantitatively, and \textcolor{black}{sometimes} even qualitatively, describe all of the relevant physical features in a variety of nonadiabatic reactive scattering and excited state relaxation processes, such as nuclear interference and detailed balance\cite{Miller2001,Kelly2016}. Hence, benchmark tests of these approaches are needed in this particular regime of the problem in order to be able to verify their viability. In order to address some of these challenges, we have recently shown the potential of the Multi-Trajectory Ehrenfest (MTEF) method to capture the correlated dynamics of a one-dimensional QED cavity-setup with a two-level atomic system coupled to a large set of cavity photon-modes\cite{HSRKA19}. Furthermore, we note that in contrast to recent work of Subotnik and co-workers, who investigated light-matter interaction with an adjusted Ehrenfest theory based method to simulate spontaneous emission of classical light \cite{CLSNS18,CLSNS218,LNSMCS18}, we focus on the description of quantized light fields.

Here we broaden our scope by investigating the performance of a comprehensive class of approximate quantum dynamics methods for simulating spontaneous emission in an optical cavity, including Ehrenfest mean-field theory\cite{Ehrenfest1927,McLachlan1964}, Tully's surface hopping algorithm\cite{Tully1990}, fully linearized \cite{Wang1998} and partially linearized \cite{Hsieh2012,fbts2} semilclassical dynamics techniques, and a selection of approximate closures for the quantum mechanical Bogoliubov-Born-Green-Kirkwood-Yvon (BBGKY) hierarchy. Through benchmark comparisons with exact numerical results, we assess the accuracy and efficiency of each method and highlight the possibilities and theoretical challenges involved with extending these approaches towards realistic systems.

The remainder of this work is divided into four sections: Sec.~\ref{Se1} gives a short overview of general quantum mechanical light-matter interactions, and a brief introduction of the class of model systems used in this study. Sec.~\ref{Se2} contains a short introduction to each of the selected dynamics methods that we consider in this work. In Sec.~\ref{Se3} we report the results of our benchmark tests of the performance of these techniques in describing spontaneous emission, stimulated absorption and strongly correlated light-matter dynamics. In Sec.~\ref{Se4} we offer some conclusions and outlooks.

\section{Electron-Photon Correlated Systems}\label{Se1}
The total Hamiltonian for a coupled light-matter system can be written as
\begin{equation}
\hat{H} = \hat{H}_{A} + \hat{H}_{F} + \hat{H}_{AF}.
\label{G9}
\end{equation}
The first term, $\hat{H}_{A}$, is the matter Hamiltonian, which may be generally expressed in the spectral representation,
\begin{equation}
\hat{H}_{A} = \sum_k \varepsilon_k | k \rangle \langle k |. \notag
\end{equation}
Here $\{\varepsilon_k,|k\rangle\}$ are the energies and stationary states of the electron-nuclei system in absence of coupling to the cavity. The second term is the Hamiltonian of the uncoupled cavity field $\hat{H}_{F}$,
\begin{equation}
\hat{H}_{F} = \frac{1}{2}\sum_{\alpha = 1}^{2N} \left(\hat{P}^{2}_{\alpha} + \omega^{2}_{\alpha}\hat{Q}_{\alpha}^{2} \right).
\label{G12}
\end{equation} The photon-field operators, $\hat{Q}_{\alpha}$ and $\hat{P}_{\alpha}$, obey the canonical commutation relation, $[\hat{Q}_{\alpha},\hat{P}_{\alpha'}] = \imath\hbar\delta_{\alpha,\alpha'}$, and can be expressed using creation and annihilation operators for each mode of the cavity field,
\begin{eqnarray}
\hat{Q}_{\alpha} = \sqrt{\frac{\hbar}{2\omega_{\alpha}}}(\hat{a}^{\dagger}_{\alpha} + \hat{a}_{\alpha}),\quad
\hat{P}_{\alpha} = i\sqrt{\frac{\hbar\omega_{\alpha}}{2}}(\hat{a}^{\dagger}_{\alpha} - \hat{a}_{\alpha}),\notag
\end{eqnarray} where $\hat{a}^\dagger_\alpha$ and $\hat{a}_\alpha$ denote the usual photon creation and annihilation operators for photon mode $\alpha$. The coordinate-like operators, $\hat{Q}_{\alpha}$, are directly proportional to the electric displacement operator, while the conjugate momenta-like operators, $\hat{P}_{\alpha}$, are related to the magnetic field \cite{Craig1998,Pellegrini2015,Flick2015}. The upper limit of the sum in Eq.~(\ref{G12}) is $2N$, as there are (in principle) two independent polarization degrees of freedom for each photon mode, however in the 1D cavity models presented here only a single polarization will be considered.

The final term in Eq.~(\ref{G9}) represents the coupling between the electron-nuclei system and the cavity field. In Coulomb gauge, and the dipole approximation \cite{Craig1998,ruggenthaler2018quantum}, this term can be written
\begin{equation} 
\label{Gint}
\hat{H}_{AF} = \sum_{\alpha=1}^{2N}\Big(\omega_{\alpha}\hat{Q}_{\alpha}(\lambda_{\alpha}\cdot \hat{\mu}) + \frac{1}{2}(\lambda_{\alpha} \cdot \hat{\mu})^2 \Big),
\end{equation} where we denote ${\hat{\mu}}$ as the electronic plus nuclear dipole moment, and ${\lambda}_{\alpha}$ as the matter-photon coupling vector \cite{Tokatly2013,ruggenthaler2014,ruggenthaler2018quantum}. The featured methodologies can be  generically applied to arbitrary complex matter systems. 

\begin{figure}[h!]{
\includegraphics[width=0.6\linewidth]{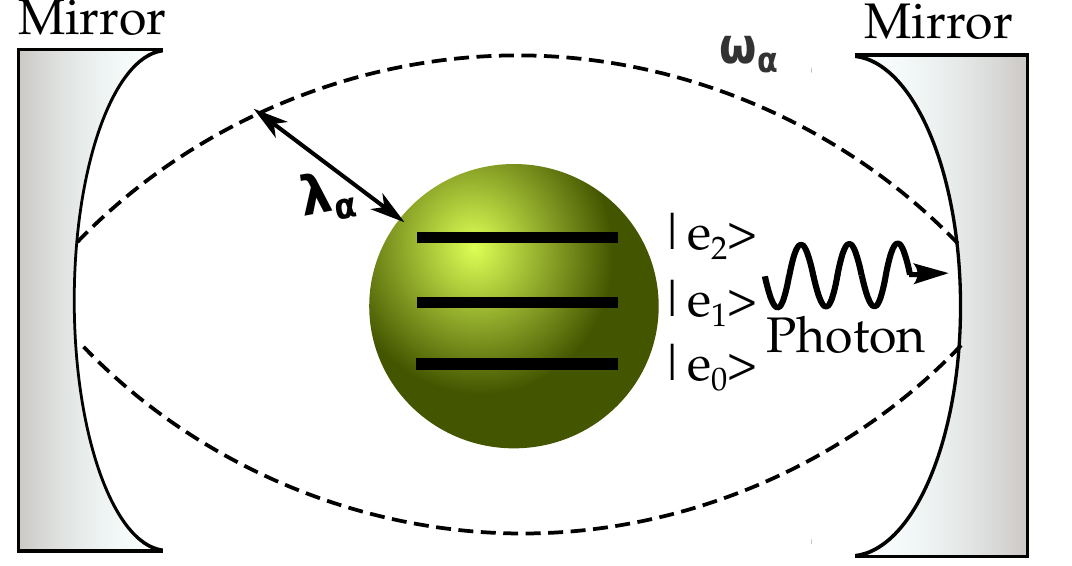}}
\caption{Cavity-setup: Few-level approximated atomic system (green) trapped in a cavity and coupled by coupling strength $\lambda_{\alpha}$ to 400 photon modes with their photonic frequency $\omega_{\alpha}$, where $\alpha = \{1,2, ...,400\}$.}
\label{AB1}
\end{figure}

With the demand for exact reference solutions, as part of the benchmarking procedure, we are forced to restrict the Hilbert-space of interest. Focusing on the evolution of the photonic degrees of freedom, we restrict the matter part to a highly simplified few-level atomic system trapped in a cavity \cite{buzek1999,Flick2017,HSRKA19} as depicted in Fig.~\ref{AB1}.
The fundamental limitations of the few-level approximation have been presented in a variety of recent publications\cite{flick2017ab,schafer2018ab,bernadrdis2018breakdown,schafer2019modification,schaefer2019rs}. While this approximation results in a strongly simplified problem, it has the advantage that exact numerical results, although nontrivial to obtain, are still achievable with a reasonable computational effort. In the case of a two-level approximation of the matter system the quadratic term $(\lambda_{\alpha} \cdot \hat{\mu})^2$ simply results in a constant energy shift and hence can be discarded \cite{schafer2018ab}. For simplicity, we also neglect this term in the case of the three level model system, \textcolor{black}{to remain consistent across set-ups and previous publications \cite{buzek1999,Flick2017,HSRKA19}.~\footnote{\textcolor{black}{We have verified that in the parameter regimes studied in this work including the quadratic term into adjusted eigenstates, according to the Hamiltonian $\hat{H}_{A} + \sum_{\alpha=1}^{2N} \frac{1}{2}(\lambda_\alpha \cdot \hat{\mathbf{\mu}})^2$, has no qualitative influence on the time-evolution of the observables associated with the cavity-bound emission process.}}
However, the quadratic term is generally important to consider as it stems from a proper definition of field observables, renders the system stable, and is essential to retain gauge and translational invariance. Applications to realistic systems should of course consider this term; for a detailed discussion of this topic, one may refer to Ref. \cite{schaefer2019rs}, for example.}

In the case of a two-level atomic system, this corresponds to a special case of the spin-boson model. With the position of the atom fixed at $r_A = \frac{L}{2}$ in this study, half of the $2N$ cavity modes decouple from the atomic system by symmetry. We adopt the same parameters as in Ref. \cite{Flick2017,Su1991}, which are based on a 1D Hydrogen atom with a soft Coulomb potential (in atomic units): $\{\varepsilon_1,\varepsilon_2\} = \{-0.6738, -0.2798\}$, $\lambda_{\alpha}(\frac{L}{2}) = 0.0103\cdot(-1)^{\alpha}$, $L = 2.362\cdot 10^{5}$ and $\mu_{12} = 1.034$. 

For the three-level atom, we adopt all the same parameters for the field and the atom-field coupling as for the two-level case. 
The atomic energies for the three level model are $\{\varepsilon_1,\varepsilon_2,\varepsilon_3\} = \{-0.6738, -0.2798, -0.1547\}$, and as before the numerical parameters are based on the 1D soft-Coulomb Hydrogen atom. The dipole moment operator only couples adjacent states, such that, the only nonzero matrix elements are $\{\mu_{12},\mu_{23}\} = \{1.034,-2.536\}$ and their conjugates. 

Furthermore, with $\frac{g_{2,1}}{\epsilon_{2} - \epsilon_{1}} = 1.2\cdot10^{-2}$ for the two-level system and $\frac{g_{3,2}}{\epsilon_{3} - \epsilon_{2}} = 2.1\cdot10^{-2}$ for the three-level system, where $g_{i,j} = \mu_{k,l}\sqrt{\frac{\epsilon_{i} - \epsilon_{j}}{2}}\lambda$ is the coupling strength for the resonant mode, our system is beyond \textcolor{black}{common perturbative approaches such as the rotating wave approximation and the well-known analytic Wigner-Weisskopf solution. The appearances of a bound photon peak in the intensity most illustratively indicates this regime. Cavity losses are not considered at this point but could be included in future developments.}

\section{Methods}\label{Se2}

\subsection{Multi-Trajectory Methods}

In this section we briefly review a selection of semiclassical dynamics methods that are based on ensembles of independent trajectories. These methods have been introduced traditionally to study electron-nuclear systems and they typically involve the use of the Wigner representation for the non-subsystem degrees of freedom. In this work we extend the application of these methods to treat coupled quantum mechanical light-matter systems, in which the degrees of freedom of the photon field will be partially Wigner transformed. The structural similarity allows for the trivial inclusion of nuclear degrees of freedom. The general expression for the average value of any observable, $\langle B (t) \rangle $, in the partial Wigner representation can be written as
\begin{eqnarray} 
\langle B (t) \rangle &=& \text{Tr}_A \int dX  \hat{B}_W(X,t) \hat{\rho}_W(X,t=0), \notag \\
 &=& \sum_{\lambda \lambda'}\int dX B^{\lambda \lambda'}_W(X,t) \rho^{\lambda' \lambda}_W(X), \notag
\end{eqnarray} where the subscript $W$ denotes the partial Wigner transform over the photonic degrees of freedom, which are represented on the continuous phase space $X=(R,P)$. \textcolor{black}{The partial Wigner transforms for an arbitrary operator $\hat{B}$ and the density matrix $\hat{\rho}$ are defined as \cite{Wigner1984}
\begin{align}
\hat{B}_{W}(R,P) &= \int dZ e^{i P \cdot Z} \langle R - \frac{Z}{2} | \hat{B} | R +\frac{Z}{2}\rangle, \notag\\
\hat{\rho}_{W}(R,P) &= \frac{1}{(2\pi\hbar)^{2N}}\int dZ e^{i P \cdot Z} \langle R - \frac{Z}{2} | \hat{\rho} | R +\frac{Z}{2}\rangle \notag.
\end{align}}
Thus, in order to assemble the average value a multi-trajectory method may be employed, which is essentially a hybrid Monte Carlo - molecular dynamics method in which initial conditions are sampled from the initial Wigner distribution, and then an ensemble of molecular dynamics trajectories is used to evaluate the time-evolution of the property of interest.

\subsubsection{Ehrenfest Mean-Field Theory}

The Ehrenfest equations of motion may be derived by assuming that the total density can be written as an uncorrelated product of the atomic and field reduced densities at all times, and then taking the appropriate classical limit\cite{Ehrenfest1927,McLachlan1964}, or by starting with the quantum-classical Liouville equation, which is formally exact for the class of systems studied here\cite{qcle}, and then making the uncorrelated approximation, i.e.,
\begin{equation}
\hat{\rho}_W(X,t)=\hat{\rho}_A(t) \rho_{F,W}(X,t), \notag
\end{equation} 
where the reduced density matrix of the atomic system is
\begin{equation}
\hat{\rho}_{A} (t) = \text{Tr}_F \Big( \hat{\rho}_W(X,t) \Big) = \int dX \hat{\rho}_W (X,t), \notag
\end{equation} and the Wigner function of the cavity field is $\rho_{F,W}(X, t) = Tr_A ( \hat{\rho}_W (X,t))$. 
The Ehrenfest mean-field equations of motion for the atomic system are:
\begin{equation}
    \partial_t \hat{\rho}_A(t) = -i\Big[ \hat{H}_A + \hat{H}_{AF,W}(X(t)), \hat{\rho}_A(t)\Big], \notag
\end{equation}
where $\hat{H}_{AF,W}$ denotes the Wigner transform of the bilinear coupling and $\hat{H}_{A}$ the atomic Hamiltonian. 
The evolution of the Wigner function of the photon field can be represented as a statistical ensemble of independent trajectories with $\mathcal{N}$ being the ensemble size, where we select uniform weights $w^j=1/\mathcal{N}$, 
\begin{equation}
\rho_{F,W} (X,t) = \frac{1}{\mathcal{N}}\sum_{j=1}^{\mathcal{N}} \delta(X-X^j(t)), \notag
\end{equation} that evolve according to Hamilton's equations of motion,
\begin{equation}
    \frac{d Q_{\alpha}}{dt} =  \frac{\partial H_{F,W}^{Eff}}{\partial P_{\alpha}},\quad \frac{d P_{\alpha}}{d t} = - \frac{\partial H_{F,W}^{Eff}}{\partial Q_{\alpha}}. \notag
\end{equation}
The mean field photonic Hamiltonian is 
\begin{equation}
    H^{Eff}_{F,W} = \frac{1}{2}\sum_{\alpha}\Big( P_{\alpha}^2 + \omega_{\alpha}^2 Q_{\alpha}^2 + 2 \omega_{\alpha} \lambda_{\alpha} Q_{\alpha}  \mu (t)   \Big), \notag
\end{equation}
where $ \mu (t)  = \text{Tr}_A(\hat{\rho}_A(0) \hat{\mu}(t))$.


\subsubsection{Fewest Switches Surface-Hopping} 
\label{sec:FBTS}
\textcolor{black}{In the following we outline the fewest switches surface hopping (FSSH) method for the electron-photon coupled system. FSSH allows feedback between the classical and quantum subsystems, however requires the photons to always propagate on a particular electronic adiabatic state, with hops between adiabatic surfaces.} \cite{T90,TP71,SS11,BR95,PR97} 

Considering the mode displacement moving along some classical trajectory $R(t) = \{R_{\alpha=1}(t),...,R_{2N}(t)\}$, the effective electronic Hamiltonian 
\begin{align*}
\hat{H}^{el}[R(t)] = \hat{H}_A + \hat{H}_{AF}[R(t)] + \frac{1}{2} \sum\limits_{\alpha=1}^{2N} \omega_\alpha^2 R_\alpha(t)^2  
\end{align*}
then becomes parametrically dependent on time through the photonic trajectory. Expanding the electronic wave function in the adiabatic basis yields
\begin{equation}
\Psi(r,R,t) = \sum_{i}c_{i}(t)\phi_{i}(r,R(t)),    \notag
\end{equation}
where $r$ denotes the collection of all electronic degrees of freedom and $c_{i}(t)$ are time-dependent complex expansion coefficients. Assuming the photonic motion with the momentum $P(t)$ to be classical, the equation of motion is given by
\begin{align*}
\partial_t \rho_{ij} &= - i \sum_{k}(H_{ik}^{el}[R(t)]\rho_{kj} - \rho_{ik}H_{kj}^{el}[R(t)]) \\ 
&- P(t) \cdot \sum_{k}(\textbf{d}_{ik}^{\alpha}[R(t)]\rho_{kj} - \rho_{ik}\mathbf{d}_{kj}^{\alpha}[R(t)] ),
\end{align*}
with the photon mode $\alpha$ and $\rho_{ij}= c_{i}(t)c^{*}_{j}(t)$ being the corresponding electronic density matrix. Furthermore, the movement of the photon is given by moving along a single potential energy surface except for some instantaneous switches. The probability for those switches, jumping from the current state $i$ to another state $j$ is defined by
\begin{equation}
g_{ij} = \frac{ b_{ij} \Delta t}{\rho_{ii}},    \notag
\end{equation}
where $\Delta t$ is a time interval from $t$ to $t+\Delta t$ and $b_{ij} = -2 \text{Re}(\rho_{ij}P(t)\cdot \textbf{d}_{ij})$ with $\textbf{d}_{ij} = \langle \phi_i(r,R(t)) \vert \partial_{R} \phi_j(r,R(t))\rangle$ being the nonadiabatic coupling vector. 

\subsubsection*{Semiclassical Mapping Methods}

Here we briefly sketch two semiclassical methods that are based on the mapping representation. These approaches can be rigorously derived from the path-integral formulation of the dynamics, or, for example, using the quantum-classical Liouville equation(QCLE)\cite{HK13}. Originally, however, the linearized semiclassical (LSC) approach has developed through a stationary-phase approximation to the full path-integral, and subsequently applying a linearization approximation to the resulting subsystem propagator\cite{Wang1998}.

With the intention of providing only the essential information about these techniques, we will briefly introduce the representation in a mapping basis, and then simply give the expressions for the corresponding equations of motion and expectation values. The interested reader may refer to specific literature (e.g. references \cite{meyermiller,stockthoss,Miller2001,StockThoss_2005,Ki08,Hsieh2012,fbts2} for example) for further information and technical details.

In order to achieve a classical-like description of the quantum subsystem, the Meyer-Miller-Stock-Thoss mapping representation\cite{meyermiller,stockthoss} is used. Each subsystem state $|\lambda\rangle$ is represented by a mapping state $|m_{\lambda}\rangle$, that is an eigenfunction of a system of $N$ fictitious harmonic oscillators, that have occupation numbers which are constrained to be 0 or 1: $|\lambda\rangle \rightarrow |m_{\lambda}\rangle$ = $|0_1, ... , 1_{\lambda},... 0_N\rangle$.

\subsubsection{Linearized Semiclassical Dynamics}\label{sec:LSC}

In the LSC method, the mapping version of an operator on the subsystem Hilbert space, $\hat{B}_{m}(X)$, is defined such that its matrix elements are equivalent to those of the corresponding  operator, $\hat{B}_{W}(X)$. For example, the mapping Hamiltonian can be written as\cite{KNK08}
\begin{equation}
\hat{B}_m(X) = \sum_{\lambda \lambda'} B_W^{\lambda \lambda'}(X) \hat{a}^{\dagger}_{\lambda}\hat{a}_{\lambda'},    \notag
\end{equation}
where the creation and annihilation operators on the subsystem mapping states, $\hat{a}^{\dagger}_{\lambda}$ and $\hat{a}_{\lambda}$,  satisfy the usual bosonic commutation relation $[\hat{a}^{\dagger}_{\lambda}, \hat{a}_{\lambda'}] = \delta_{\lambda \lambda'}$. Completing the Wigner transform over the subsystem, the mapping Hamiltonian can be written as a function of continuous phase space variables $(X,x) = (R,P,r,p)$,
\begin{equation}
B_m(X) = \sum_{\lambda \lambda'} B_W^{\lambda \lambda'}(X) (r_{\lambda} r_{\lambda'}+ p_{\lambda} p_{\lambda'}-\delta_{\lambda \lambda'}).  \notag 
\end{equation}

The LSC time-evolution of an arbitrary operator in the mapping representation, $B_m(X)$, can be written as a classical-like dynamics in the extended Wigner-mapping phase space,
\begin{equation}
\frac{\partial}{\partial t}B_m (X,x,t) = \big\{ H_m(X,x), B_m(X,x,t) \big\}_{{X,x}}. \notag
\end{equation}
Due to the Poisson bracket structure of this equation the density can be obtained from the evolution of an ensemble of independent trajectories, $\rho_m(X,t)={\mathcal N}^{-1}\sum_{i=1}^{{\mathcal N}} \delta(X-X_i(t))$, where the $X_i(t)=(R_i(t),P_i(t))$ are given by the solutions of the following set of ordinary differential equations \cite{NBK10}:
\begin{eqnarray}
\frac{d r_{\lambda}}{dt}&=& \frac{\partial H_m}{\partial p_\lambda}, \qquad
\frac{d p_{\lambda}}{dt}=-\frac{\partial H_m}{\partial r_{\lambda}}, \notag\\
\frac{d R}{dt} &=& \frac{\partial H_m}{\partial  P}, \qquad
\frac{d P}{dt}= -\frac{\partial H_m}{\partial R}.\nonumber
\end{eqnarray}

\subsubsection{Partially Linearized Quantum - Classical Dynamics}

A less severe approximation to the QCLE\cite{Huo2011,Hsieh2012} uses a partially linearized approximation to the equations of motion for the coupled system, using the mapping representation for the forward and backward time-propagators separately. This doubles the number of mapping variables used to describe each subsystem state, but yields an efficient approximate solution to the QCLE in this forward-backward mapping form. This forward-backward trajectory solution (FBTS) describes a classical-like dynamics in the extended phase space of the environmental and the mapping variables that represent the subsystem degrees of freedom. The effective Hamiltonian function that generates the FBTS evolution is
\begin{equation}
H_e(X,x,x') = \frac{1}{2}(H_m(X,x)+H_m(X,x'))	    \notag
\end{equation} where $(X,x,x') = (R,P,r,r',p,p')$. 

The continuous trajectories that define the FBTS solution to the quantum-classical Liouville equation can be represented by the following Hamiltonian equations of motion \cite{KZSK12},
\begin{align*}
&\frac{dr_{\mu}}{dt} = \frac{\partial H_{e}(X,x)}{\partial p_{\mu}} , ~~\quad \frac{dp_{\mu}}{dt} = -\frac{\partial H_{m}(X,x)}{\partial r_{\mu}},\nonumber \\
&\frac{dr'_{\mu}}{dt} = \frac{\partial H_{m}(X,x')}{\partial p'_{\mu}} , \quad \frac{dp'_{\mu}}{dt} = -\frac{\partial H_{m}(X,x')}{\partial r'_{\mu}},\\
&\frac{d R}{dt} = \frac{P}{M}, ~\quad\qquad\qquad \frac{d P}{dt} = -\frac{\partial H_{e}(X,x,x')}{\partial R}.\nonumber
\end{align*}

In the FBTS simulation algorithm, the matrix elements of the operator $\hat{B}_W(t)$ are approximated using the following expression,
\begin{align*}\nonumber 
B^{\lambda \lambda'}_W(X,t) =& \sum_{\mu \mu} \int dx dx' \phi(x) \phi(x') (r_{\lambda} + i p_{\lambda})(r_{\lambda}' - i p_{\lambda}') \\ \times B^{\mu \mu'}_W&(X_t) (r_{\mu}(t) + i p_{\mu}(t))(r_{\mu'}'(t) - i p_{\mu'}'(t)),
\end{align*}
where $\phi(x) = e^{-\sum_{mu}(r_{\mu}^2+p_{\mu}^2)}$ are normalised Gaussian distribution functions, and evaluation of the integrals over the time-independent $\phi(x)$ functions is carried out by Monte Carlo sampling. 


\subsection{Quantum BBGKY-Hierachy}
In the following we briefly describe the quantum mechanical BBGKY-hierarchy, which is an exact reformulation of many-body quantum dynamics. As such it can capture quantum interference and fluctuations. In practice, some approximate closures for the hierarchy have to be employed to reduce the computational cost of this approach. For a system of interacting fermions and bosons according to Eq.~\eqref{Gint}, where we focus on the explicit Pauli-spin representation of the 2-level system, i.e.,
\begin{align}
\label{eq:ham3}
\hat{H} &=-\frac{\Delta\varepsilon}{2} \hat{\sigma}_z + \frac{1}{2}\sum_{\alpha=1}^{2N}\Big(\hat{P}^{2}_{\alpha}+ \omega^{2}_{\alpha}\hat{Q}^{2}_{\alpha}\Big) + \hat{E}(r_A)\hat{\sigma}_x,\\
&\hat{E}(r_A) = \sum_{\alpha=1}^{2N}\mu_{12}\omega_{\alpha}\lambda_{\alpha}(r_A)\hat{Q}_{\alpha},\notag \end{align}
with $\Delta\varepsilon=\varepsilon_2-\varepsilon_1$, the underlying equations of motion, known as the quantum BBGKY-hierarchy~\cite{ShunJin1985,Fricke1996,Bonitz2016} follow from the Heisenberg equations of motion for the Hamiltonian. Consistent with previous publications\cite{sakkinen2014,sakkinen2015}, we introduce the short-hand notation $\hat{X}_{1\alpha}\equiv\hat{Q}_{\alpha}$, $\hat{X}_{2\alpha}\equiv\hat{P}_{\alpha}$ such that the correlation functions are given by

\begin{align*}
   \Lambda_{i\alpha,j\beta}
   &\equiv\langle\hat{X}_{i\alpha}\hat{X}_{j\beta}\rangle
   -X_{i\alpha}X_{j\beta}
   \, ,\\
   \Lambda_{\varepsilon;j\alpha}
   &\equiv\langle\hat{X}_{j\alpha}\hat{\sigma}_{\varepsilon}\rangle
   -X_{j\alpha}\sigma_{\varepsilon},
   \, 
\end{align*}

with \textcolor{black}{ $i,j \in \{1,2\},~\varepsilon \in \{x,y,z\}$} and we chose to suppress the 
time-arguments for brevity. In this work we truncate the infinite hierarchy of equations of motion at the doublets level for the correlation
functions \cite{Hoyer2004}, resulting in an approximation conventionally referred to as the second Born approximation \cite{Zimmermann1994,Lohmeyer2005}. 
This extends the Hartree-Fock-type approximation as presented in \cite{Pellegrini2015,Flick2017} to the next higher consistent approximation level of the hierarchy.
With $\textbf{X}\equiv  (Q_{\alpha=1},\dots,Q_{\alpha=(2N)},P_{\alpha=1},\dots,P_{\alpha=(2N)})^{T} = (X_{11},\dots,X_{1(2N)},X_{21},\dots,X_{2(2N)})^{T}$ the normal coordinate averages satisfy
\begin{align*}
   \dot{\textbf{X}}
   &=\big\{\textbf{X}, H_{\mathrm{cl}}(\sigma_{x},\textbf{X})\big\}
   \, ,
\end{align*}

where $\{\cdot,\cdot\}$ denotes the canonical Poisson bracket. Furthermore, $H_{\mathrm{cl}}$ defines the classical Hamiltonian function, i.e., providing the classical equivalent to Eq.~\eqref{eq:ham3} $\hat{B} \rightarrow \langle B \rangle$.
The spin-projection averages in turn obey the equations
\begin{align*}
   \dot{\sigma}_{z}
   &=2E(r_A)\sigma_{y}
   +2\boldsymbol{\lambda}_{\mathrm{eff}}^T\cdot \boldsymbol{\Lambda}_{y}
   \, ,\\
   \dot{\sigma}_{y}
   &=-\Delta\varepsilon\sigma_{x}
   -2E(r_A)\sigma_{z}
   -2\boldsymbol{\lambda}_{\mathrm{eff}}^T\cdot\boldsymbol{\Lambda}_{z}
   \, ,\\
   \dot{\sigma}_{x}
   &=\Delta\varepsilon\sigma_{y}
   \, ,
\end{align*}

where $\boldsymbol{\lambda}_{\mathrm{eff}}\equiv(\omega_{1}\lambda_{1}(r_A)\mu_{\mathrm{12}},
\dots,\omega_{M}\lambda_{(2N)}(r_A)\mu_{\mathrm{12}})^{T}$ represents 
the effective light-matter coupling. Moreover, we introduced the vector notation
$\boldsymbol{\Lambda}_{\varepsilon}\equiv(\Lambda_{\varepsilon;11},\dots,
\Lambda_{\varepsilon;1(2N)},\Lambda_{\varepsilon;21},\dots,
\Lambda_{\varepsilon;2(2N)})^{T}$ for the correlation functions. The dynamics of
the correlation functions are determined by

\begin{align*}
   \dot{\boldsymbol{\Lambda}}_{z}
   &=\big\{\boldsymbol{\Lambda}_{z},
   H_{\mathrm{cl}}(-\imath\sigma_{y}-\sigma_{x}\sigma_{z},\boldsymbol{\Lambda}_{z})\big\}
   \notag\\
   &\phantom{=}
   +2E\boldsymbol{\Lambda}_{y}
   +2\sigma_{y}\boldsymbol{\Lambda}\cdot \boldsymbol{\lambda}_{\mathrm{eff}}(r_A)
   \, ,\\
   \dot{\boldsymbol{\Lambda}}_{y}
   &=\big\{\boldsymbol{\Lambda}_{y},
   H_{\mathrm{cl}}(-\imath\sigma_{z}-\sigma_{y}\sigma_{x},-\boldsymbol{\Lambda}_{y})\big\}
   \notag\\
   &\phantom{=}
   -\Delta\varepsilon\boldsymbol{\Lambda}_{x}
   +2E\boldsymbol{\Lambda}_{z}
   +2\sigma_{z}\boldsymbol{\Lambda}\cdot\boldsymbol{\lambda}_{\mathrm{eff}}(r_A)
   \, ,\\
   \dot{\boldsymbol{\Lambda}}_{x}
   &=\big\{\boldsymbol{\Lambda}_{x},H_{\mathrm{cl}}(1-\sigma_{x}^{2},\boldsymbol{\Lambda}_{x})\big\}
   \notag\\
   &\phantom{=}
   -\Delta\varepsilon\boldsymbol{\Lambda}_{y}
   \, ,
\end{align*}

where the matrix $\boldsymbol{\Lambda}$ with the elements $\Lambda_{i\alpha,j\beta}$ is the 
covariance matrix satisfying the equation

\begin{align*}
   \dot{\boldsymbol{\Lambda}}
   &=\boldsymbol{J}\cdot\boldsymbol{\Omega}\cdot\boldsymbol{\Lambda}
   -\boldsymbol{\Lambda}\cdot\boldsymbol{\Omega} \cdot\boldsymbol{J}
   -\boldsymbol{\lambda}_{\mathrm{eff}}\cdot\boldsymbol{\Lambda}_{x}^{T}
   -\boldsymbol{\Lambda}_{x}\cdot\boldsymbol{\lambda}_{\mathrm{eff}}^{T}
   \, .
\end{align*}

Here $\boldsymbol{J}$ is the standard symplectic matrix 
\begin{align*}
\boldsymbol{J} =  
\begin{pmatrix} 
0 & 1 & 0 &  \\
-1 & 0 & 1 & \dots \\
0 & -1 & 0 &  \\
&\dots
\end{pmatrix}
\end{align*}
and $\boldsymbol{\Omega}$ denotes a matrix such 
that $\Omega_{1\alpha,1\alpha}=\omega_{\alpha}^{2}$,
$\Omega_{2\alpha,2\alpha}=1$, and otherwise zero.\\
Evolving the covariance matrix in time allows the field fluctuations to dynamically respond to the polarizable matter. Deriving the equation of motions from the many-body perturbation hierarchy sets an implicit condition on the dynamic fluctuations as the 2-particle reduced density matrix has to be identically zero to guarantee that only a single electron is acting in our system. In the following section we will show that enforcing this condition cures almost completely all nonphysical negative intensities that arise otherwise and overall improves the performance of the second Born approximation considerably.

\subsection{\textcolor{black}{Configuration Interaction Expansion}}\label{sec:ci}
\textcolor{black}{
To obtain accurate reference solutions, considered as exact benchmarks for this low dimensional model, we truncate the Configuration Interaction (CI) expansion such that we allow at most two photons per mode, featuring 400 modes, while retaining the full two and three state representation for the atomic system.}
\begin{align}
\label{eq:ci}
\begin{split}
\vert \Psi (t) \rangle &= \sum\limits_{k}  c_{k,0}(t) \vert k\rangle \otimes \vert 0\rangle\\
&+ \sum\limits_{k}\sum\limits_{n_1}^{2N}  c_{k,n_1}(t) \vert k\rangle \otimes \hat{a}^\dagger_{n_1}\vert 0\rangle\\
&+ \sum\limits_{k}\sum\limits_{n_1,n_2}^{2N^2+N}  c_{k,n_1,n_2}(t) \vert k\rangle \otimes \hat{a}^\dagger_{n_1}\hat{a}^\dagger_{n_2}\vert 0\rangle
\end{split}
\end{align}
\textcolor{black}{In line with the nature of CI expansions, the numerical cost exponentially grows when increasing the number of allowed photonic excitations. When exploiting the bosonic symmetry of the photons in total $1 + 2N + 2N(2N-1)/2$ photon basis functions span the zero-photon (vacuum), one-photon (1pt) and two-photon (2pt) space. Combined with the low-dimensional matter system featuring the eigenstates $\vert k \rangle$, it is computationally non-trivial but feasible to propagate this CI expanded wavefunction using the Lanczos algorithm \cite{park1986unitary,flick2016exact}. We ensured that the above (vacuum+1pt+2pt) CI basis is sufficient for the observables and parameters studied in this work.\footnote{ \textcolor{black}{As the exponential scaling permits the inclusion of higher photon states for the given model, we ensured convergence investigating a related 3-level system based on a screened Hydrogen atom with $1/10$ of the atomic binding potential coupled to the 100 lowest harmonics of the former cavity. Including the three-photon states resulted in marginal numerical changes such that we deem the selected two-photon states sufficient for the investigated model.} } Although spontaneous decay from the 2-level atomic system will lead to at most a single observable photon, the photonic fluctuations can reach the 2pt state space which results in the possibility to bind photon intensity at the atomic position (see Fig.~\ref{ABNew}). 
}

\section{Results and Discussion}\label{Se3}

As in earlier work\cite{HSRKA19}, we note that the Wick normal ordered form for operators (denoted $:\hat{B}:$ for some operator $\hat{B}$) is used when calculating average values in this study. The reason for using the normal ordered form, in practice, is to remove the  typically non-measurable \cite{riek2015direct, benea2019electric} effect of vacuum fluctuations from the results, which ensures that both $\langle E \rangle = 0$ and $\langle I \rangle = 0$, irrespective of the number of photon modes in the cavity field, when the field is in the vacuum state. \textcolor{black}{ In order to guarantee a distinct spacial resolution for the dynamics of the photonic wavepacket in the cavity and to ensure the inclusion of all possible inference effects we use 400 photon modes to represent the cavity field that is coupled to a two or three energy-level atomic system in all calculations shown below.} We choose the atom to be initially in the highest excited state, and the cavity field in the vacuum state at zero temperature. For our benchmark numerical treatment we solved the time-dependent Schr\"odinger equation by using a truncated Configuration Interaction expansion as introduced in Sec.~\ref{sec:ci}. \textcolor{black}{The atomic population operator is given by 
$\hat{\sigma}_{i}(t) = |c_{i}(t)|^2 ,$
where $c_{i}(t)$ denotes the time-dependent CI coefficient for the corresponding atomic energy level. Furthermore we define the normal-ordered electric field intensity operator as
\begin{equation}
:\hat{I}(r,t): = :\hat{E}^2(r,t): = 2\sum_{\alpha=1}^{2N} \omega_{\alpha}\zeta^{2}_{\alpha}(r)\hat{Q}_{\alpha}^{2}(t) - \sum_{\alpha=1}^{2N}\zeta^{2}_{\alpha}(r). \nonumber
\label{G15}
    \end{equation}
    with 
\begin{equation}
\zeta_{\alpha}(r) = \sqrt{\frac{ \omega_{\alpha}}{\epsilon_{0} L}} \sin\Big(\frac{\alpha \pi}{L} r\Big). \nonumber
\end{equation}}

\subsection{2-Level Atom: One-Photon Emission Process}

\begin{figure}[h!]{
\includegraphics[width=1\linewidth]{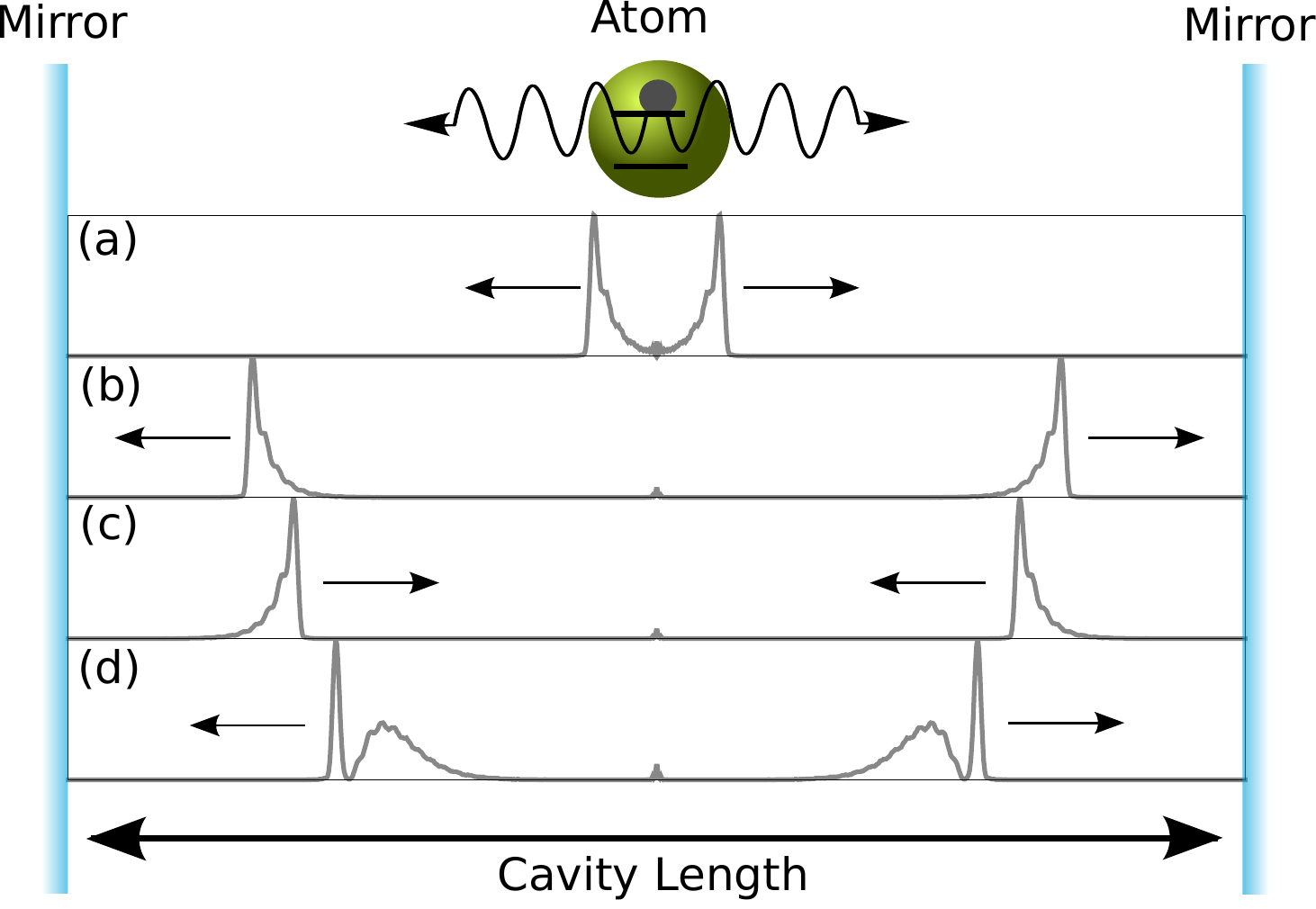}}
\caption{A schematic sketch of the photon-field intensity propagating through the cavity for four time snap-shots: (a) $t=100~a.u.$, (b) $t=600~a.u.$, (c) $t=1200~a.u.$, (d) $t=2100~a.u.$.}
\label{overview}
\end{figure}

In Fig.~\ref{overview} we show a schematic sketch of the propagating photon-field intensity along the axis of the cavity for four different time snap-shots. As the spontaneous emission process evolves, a photon wave-packet with a sharp front is emitted from the atom (e.g. panel (a) of Fig.~\ref{overview}) and travels towards the boundaries (e.g. panel (b) of Figs.~\ref{overview}) where it is reflected, and then travels back to the atom (e.g. panel (c) of Fig.~\ref{overview}). The emitted photon is then absorbed and re-emitted by the atom, which results in the emergence of interference phenomena in the electric field. This produces a photonic wave packet with a more complex shape (e.g. panel (d) of Fig.~\ref{overview}). In Figs.~\ref{Appendix1}, and \ref{AB3} we plot this spontaneous emission process for the different methods compared to the exact result (black dashed line). Here we observe that the essential differences among the methods are (i) determining the correct amplitude of the wave-packet, (ii) capturing the re-emission interference pattern and (iii) resembling the bound photon at the atomic position.

\subsubsection{Finite size corrections to the BBGKY hierarchy}
\label{SecApp1}

\begin{figure}[h!]{
\includegraphics[width=0.8\linewidth]{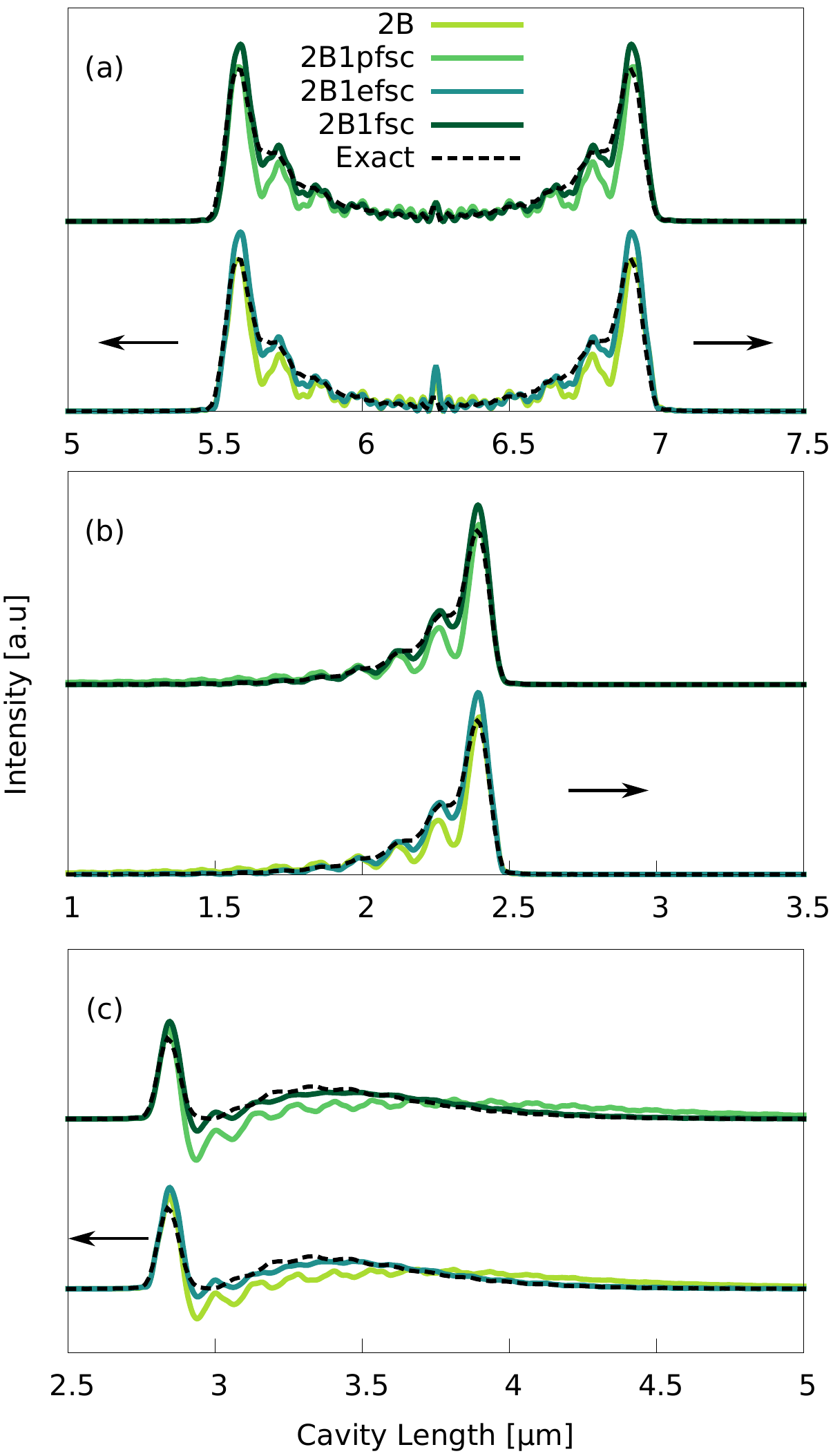}}
\caption{Intensity of the emitted (normal-ordered) photon field using different finite-size corrections at  three different time snapshots: (a) $t = 100 ~a.u.$, (b) $t = 1200 ~a.u.$, (c) $t = 2100 ~a.u.$; no correction, single-photon correction (1pfsc), single-electron correction (1efsc) and single-photon and single-electron correction (1fsc) for the BBGKY hierarchy within the second Born approximation. The arrow indicates the direction of the wave packet.}
\label{Appendix1}
\end{figure}

By partially summing the infinite series of perturbative diagrams that arise as a consequence of the Heisenberg equation of motion \textcolor{black}{using Hamiltonian \eqref{eq:ham3}}, we intrinsically introduce spurious interaction between physically non-existent particles as we consider more diagrams than particles are present in the physical system. This is a well-known subject of interest in electronic structure theory \cite{kremp1997non,von2009successes,verdozzi2011some,stefanucci2013nonequilibrium,florian2013equation,leymann2014expectation,richter2009few}. Specifically for our problem, this can result in such fundamental violations as producing negative atomic state occupations or photon field intensities (see Fig.~\ref{Appendix1}). Enforcing the correct fermionic truncation of the many-body hierarchy acts to cure most of the nonphysical features that appear, i.e., negative intensities after the re-emission and strong oscillations around the exact solution. This restriction to the single electron subspace (1efsc) is performed by enforcing that the two-particle reduced density matrix be identically zero, $\rho^{(2)}_{ijkl} = 0$. For one-body reduced density-matrices $\rho^{(1)}_{ij}$, the cluster expansion on the exchange-only level $\rho^{(2)}_{ijkl} \approx \rho^{(1)}_{il} \rho^{(1)}_{jk} - \rho^{(1)}_{ik} \rho^{(1)}_{jl}$ guarantees this if $\rho^{(1)}_{ij}$ is idempotent.

A further correction is possible in the photonic subspace, i.e., enforcing at most a single photon in the cavity for the two-level system (1pfsc). This is achieved by substituting higher correlation matrices with lower order expansions such that the equation of motions do not connect to higher excitations and corrects the bound photon intensity to excellent accuracy. Employing both restrictions at the same time (1fsc) leads to the overall best performance and we focus on those results in Sec.~\ref{Se3}. For multiple electrons and photonic excitations such corrections will become less relevant and less straightforward to apply.

\subsubsection{Trajectory-based Semiclassical methods}

\begin{figure}[h!]{
\includegraphics[width=1\linewidth]{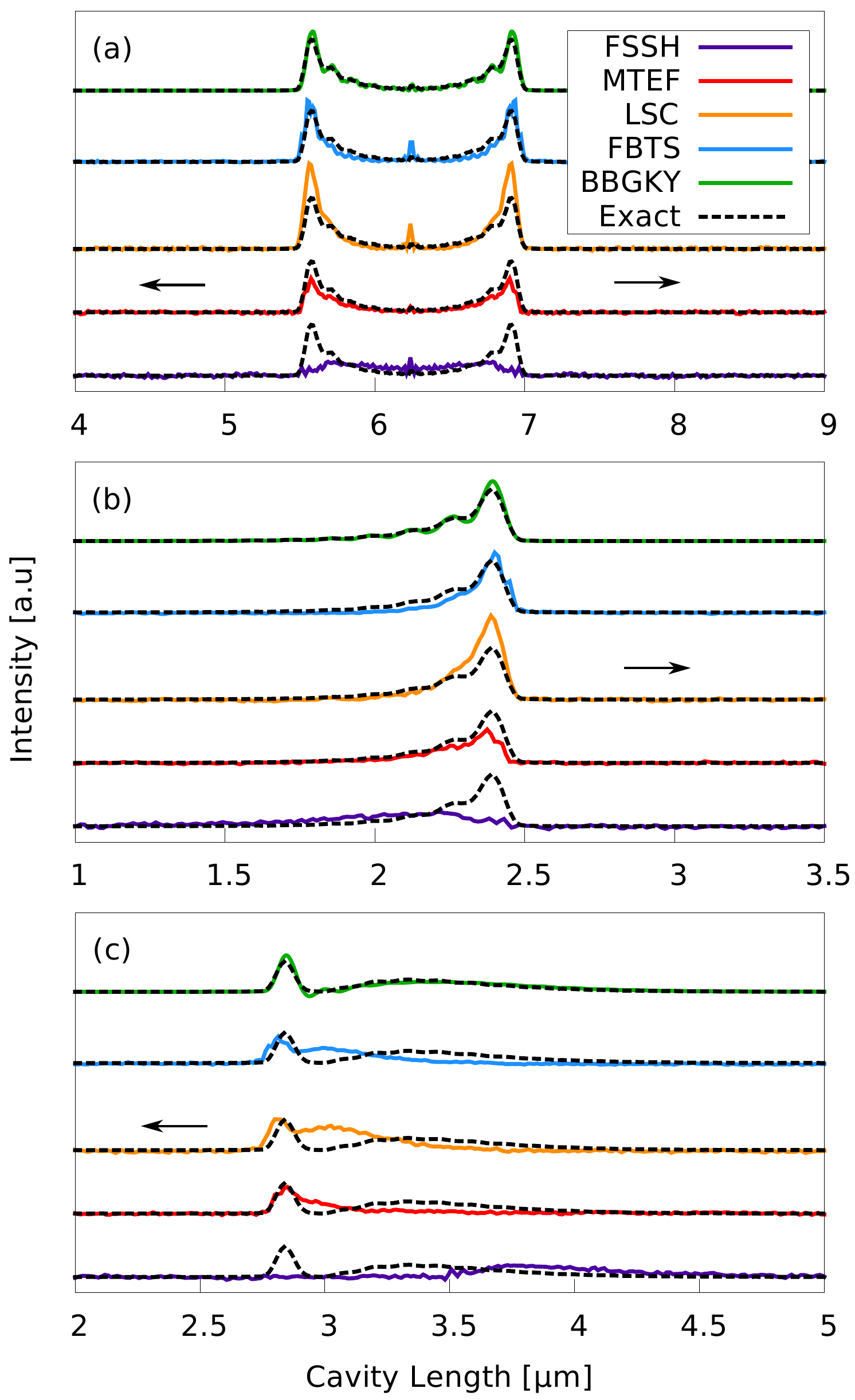}}
\caption{Time-evolution of the average field intensity for the one-photon emission process, at  three different time snapshots: (a) $t = 100 ~a.u.$, (b) $t = 1200 ~a.u.$, (c) $t = 2100 ~a.u.$. Exact solution (black-dashed), FSSH (purple), MTEF (red), LSC (orange), FBTS (blue) and (1fsc)BBGKY (green). The arrow indicates the direction of the wave packet.}
\label{AB3}
\end{figure}

To perform numerical simulations using the semiclassical dynamics methods, we first employ Monte Carlo sampling from the Wigner transform of the initial density operator of the photon field, $\hat{\rho}_{F,W}(X,0)$, to generate an ensemble of initial conditions for the trajectory ensemble $(Q_{\alpha}^j(0),P_{\alpha}^j(0))$. The Wigner transform of the zero temperature vacuum state is given by 
\begin{equation}
    \rho_{F,W}(X,0)=\prod_{\alpha=1}^{2N} \frac{1}{\pi}\exp{\left[-\frac{P^{2}_{\alpha}}{\omega_{\alpha}} - \omega_{\alpha} Q_{\alpha}^2\right]}. \notag
\end{equation} 

We then evolve each initial condition independently according to the corresponding equations of motion to produce a trajectory. Average values are then constructed by summing over the entire trajectory ensemble, and normalizing the result with respect to $\mathcal{N}$, the total number of trajectories. We use an ensemble of $\mathcal{N} = 10^5$ independent trajectories for the MTEF, FSSH, LSC, and FBTS calculations, sampled from the Wigner transform of the initial field density operator. This level of sampling is sufficient to converge the atomic observables to graphical accuracy, while the field intensity would require a slightly larger trajectory ensemble for graphical convergence. 

\begin{figure}[h!]{
\includegraphics[width=0.5\textwidth]{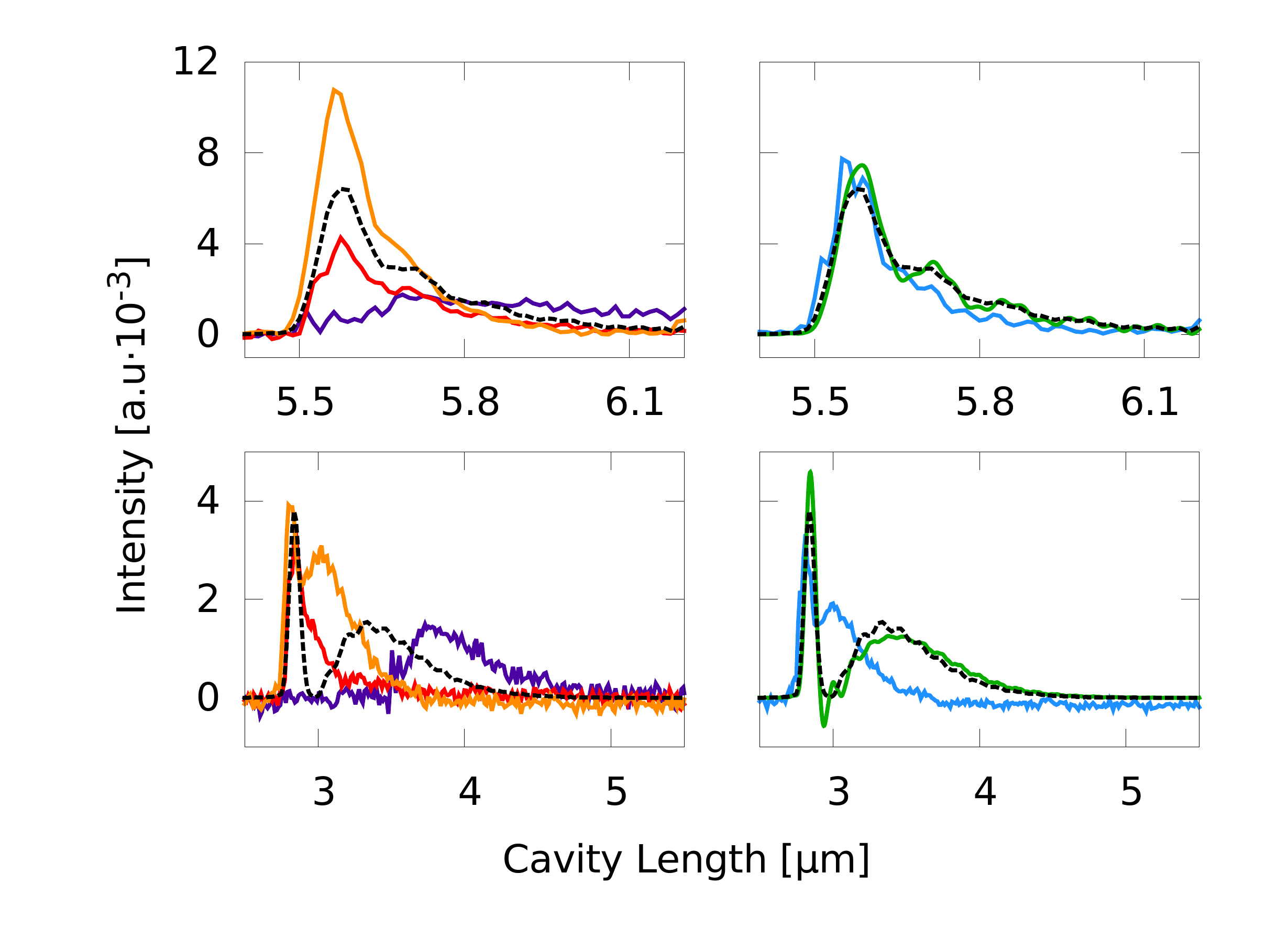}}
\caption{Zoom-in onto the wavefronts of Fig.~\ref{AB3} (same color code) at time $t = 100~a.u.$ (upper panel) and $t = 2100~a.u.$ (lower panel).}
\label{AB4}
\end{figure}

\begin{figure}[h!]{
\includegraphics[width=1\linewidth]{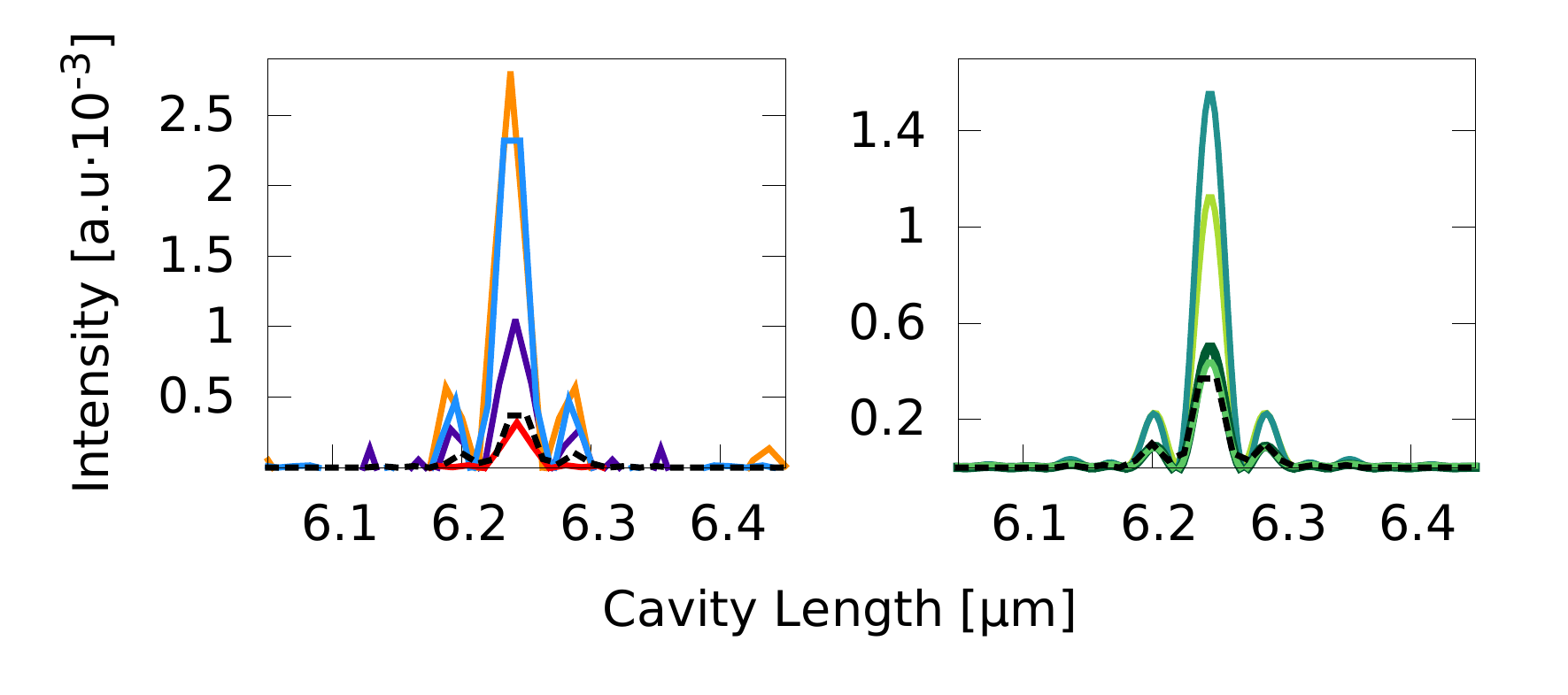}}
\caption{Left: Zoom-in on the bound photon state of Fig.~\ref{AB3} (same color code). Right: Zoom-in on the bound photon state of Fig.~\ref{Appendix1} (same color code).}
\label{AB6}
\end{figure}

In order to illustrate the comparison more accurately a zoom-in of Fig.~\ref{AB3} is depicted in Fig.~\ref{AB4}, and \ref{AB6} in the same coloring. We find that the shapes of the (2B-1fsc) BBGKY-method and the FBTS-method nicely agree with the exact wave-packet shape for time $100$ [a.u.], while the MTEF and LSC simulations are qualitatively accurate, but miss the correct wave-packet amplitude. We find that FSSH performs rather poorly, as it fails to capture the qualitative structure of the outgoing wave-packet. Further, we observe at time $2100$ [a.u.] that the FSSH-method has broken-down completely as it fails to reproduce the wave-packet structure in addition to exhibiting a time-delay. Considering the other trajectory-based methods, we find that MTEF is not able to reproduce the photon re-emission due to the lack of capturing interferences within mean-field methods. On the other hand FBTS and LSC predict a substantial amount of interference in the form of a second maximum, however shifted to earlier times in relation to the exact solution. As seen previously, the corrected second Born truncation of the BBGKY hierarchy is in very good agreement with the exact simulation; nevertheless it still develops very small unphysical negative intensity values in between the first and second wave-packet maxima.

All methods are capable of describing the remaining intensity at the atomic position. This intensity corresponds to the bound photon intensity, which emerges from beyond rotating-wave approximation (RWA) effects. \textcolor{black}{More precisely, in Fig. \ref{ABNew} we show the photon field intensity for the exact reference solution calculated in four different ways according to Eq.~\eqref{eq:ci}. First including all two-photon states (2pt) without RWA (blue) and then performing the same calculation within RWA (cyan). Here we find that using the RWA erases the bound photon state. Furthermore, we find that only including the one-photon states (1pt) is also not sufficient to capture this higher-order effect, as in both cases without RWA (red) and with RWA (orange) no bound photon is observed.
\begin{figure}[h!]{
\includegraphics[width=0.5\textwidth]{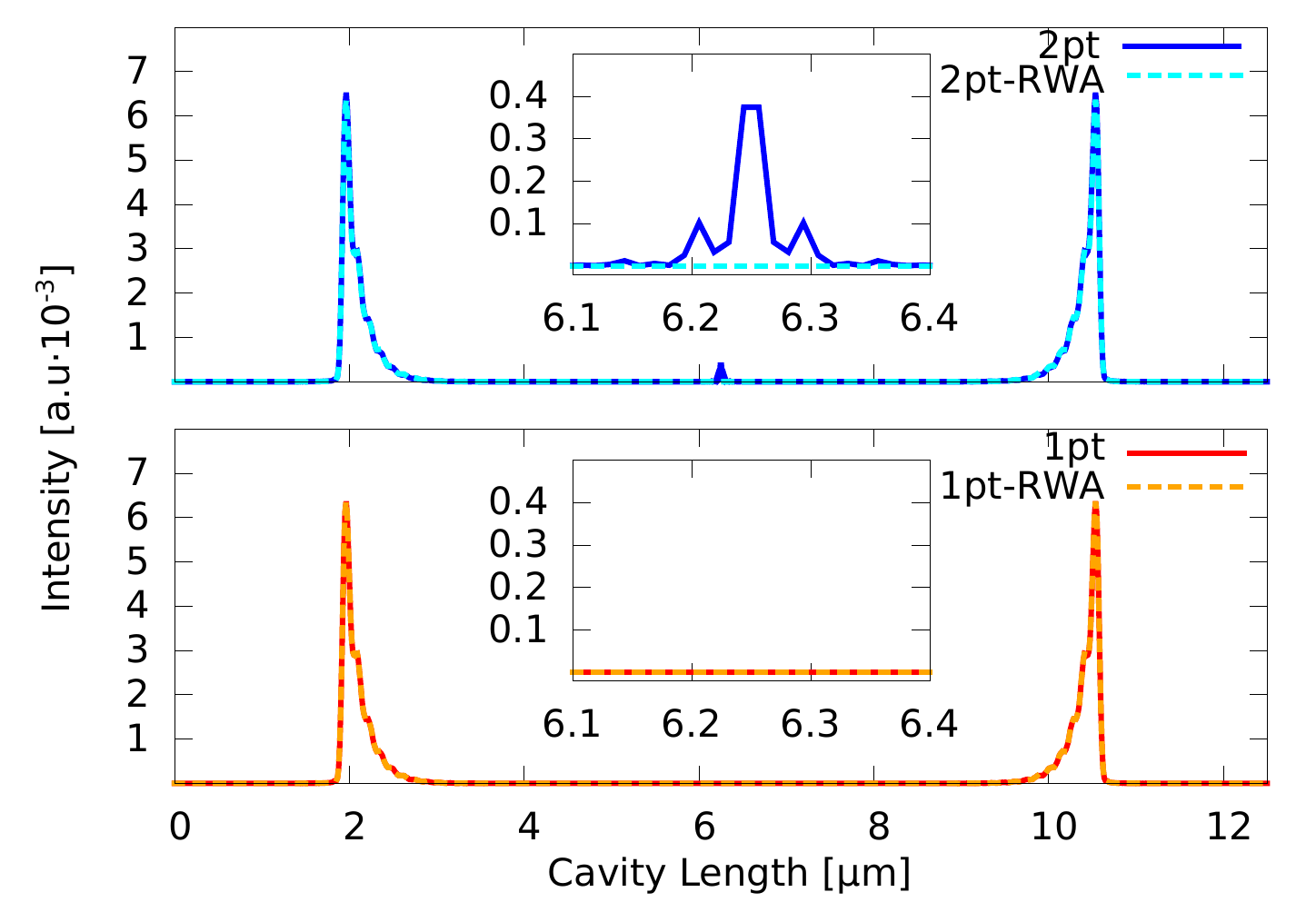}}
\caption{Photon field intensity for the exact reference solution at time $600$[a.u.] for Blue: including two-photon states (2pt) and no RWA, Cyan: including two-photon states (2pt) with RWA, Red: including only one-photon states (1pt) and no RWA, Orange: including only one-photon states (1pt) with RWA.}
\label{ABNew}
\end{figure}
Therefore, those results show that all methods are indeed capable of describing effects beyond the perturbative regime such as bound photon states.} In Fig.~\ref{AB6} we depict this signature feature of the bound photon state for time $1200$ [a.u.]. Here we find, that BBGKY and MTEF perform best, as FBTS, LSC and FSSH overestimate the amplitude for the remaining intensity. Without single photon correction the BBGKY amplitude is comparable to the one of FBTS, i.e., finite size corrections in both, fermionic and photonic subspace, are important to obtain excellent results.

\begin{figure}[h!]{
\includegraphics[width=1.1\linewidth]{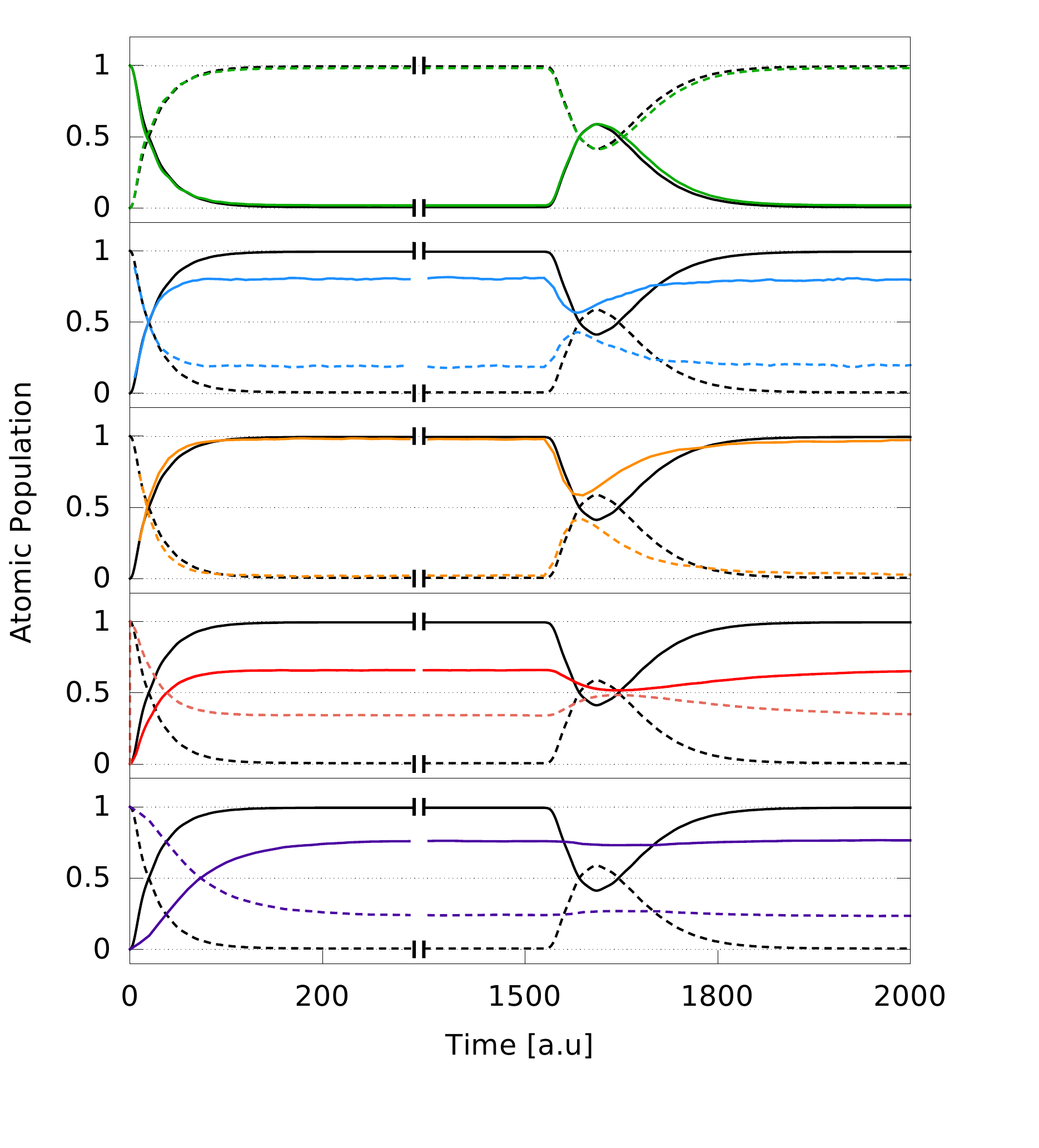}}
\caption{Time-evolution of the atomic state population in the same color code as Fig.~\ref{AB3}. Solid lines represent the atomic ground state, and dashed lines represent the excited state.} 
\label{AB5}
\end{figure}

\textcolor{black}{In Fig.~\ref{AB5} we plot the atomic adiabatic state population in the same color code as in Fig.~\ref{AB3}.} Here BBGKY leads to excellent accuracy while among the trajectory methods LSC performs best. The initial decay, which is connected to the shape of the wave-front, is however superior in FBTS with the drawback of an incomplete de-excitation. While MTEF is capable of qualitatively describing the process, it fails on quantitative scales
and even worse is FSSH which not even qualitatively resembles the process.

\subsection{3-Level Atom: Two-Photon Emission Process}

Let us turn our attention to the slightly more complex three-level system where we focus on the most promising approaches with respect to extrapolations towards realistic systems in mind. We thus exclude FSSH due to its relatively poor performance and BBGKY due to its high computational effort, which we will later discuss in more detail.

\begin{figure}[h!]{
\includegraphics[width=1\linewidth]{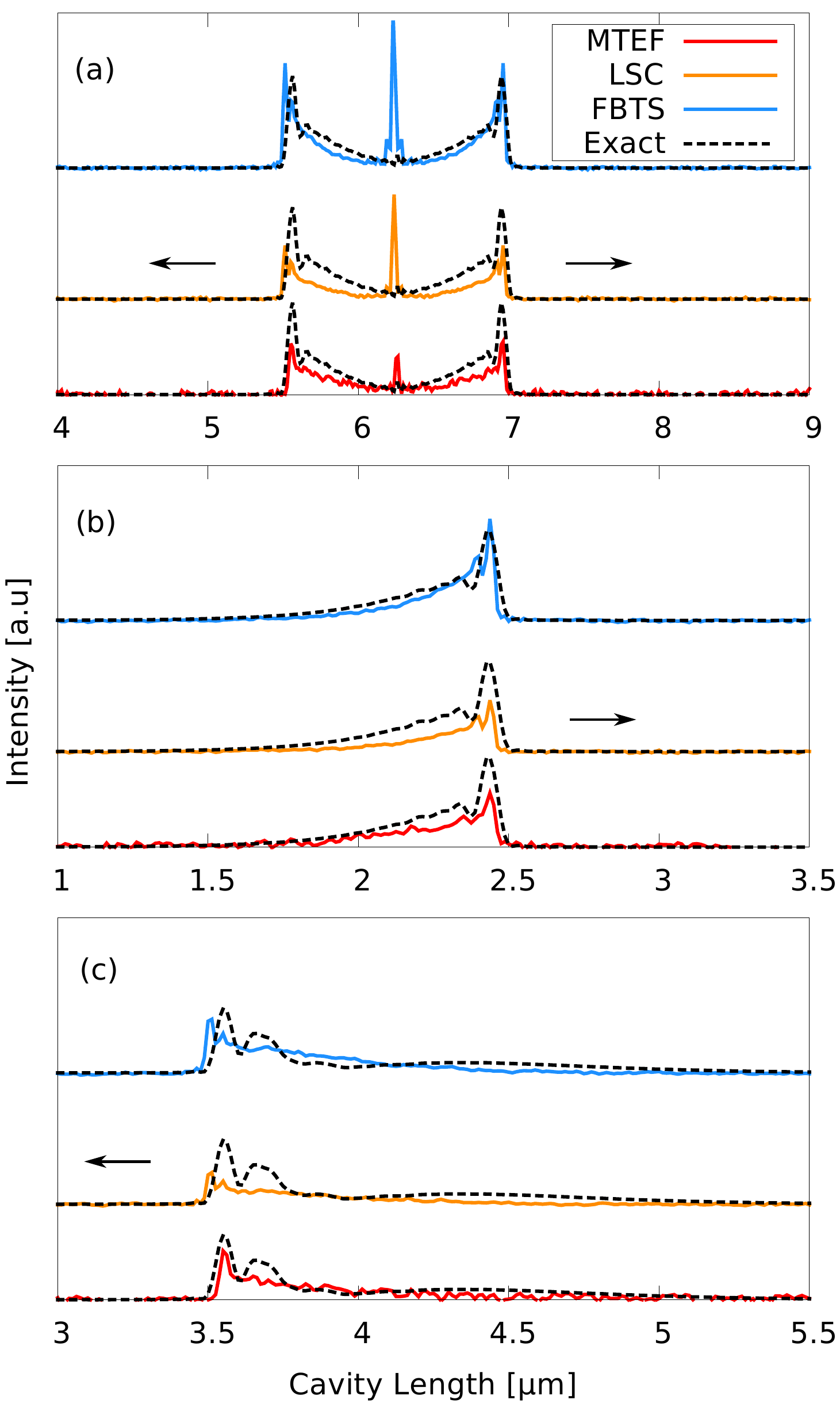}}
\caption{Time-evolution of the average field intensity for the two-photon emission process, at three different time snapshots: (a) $t = 100 ~a.u.$, (b) $t = 1200 ~a.u.$, (c) $t = 2100 ~a.u.$. Exact solution (black-dashed), MTEF (red), LSC (orange) and FBTS (blue). Please note that in this plot the amplitude of the bound photon state for the FBTS simulation is reduced in order to improve the illustration of the results. Explicit quantitative results for the bound photon state can be found in Fig.~\ref{AB9}. The arrow indicates the direction of the wave packet.  }
\label{AB7}
\end{figure}
\begin{figure}[h!]{
\includegraphics[width=0.5\textwidth]{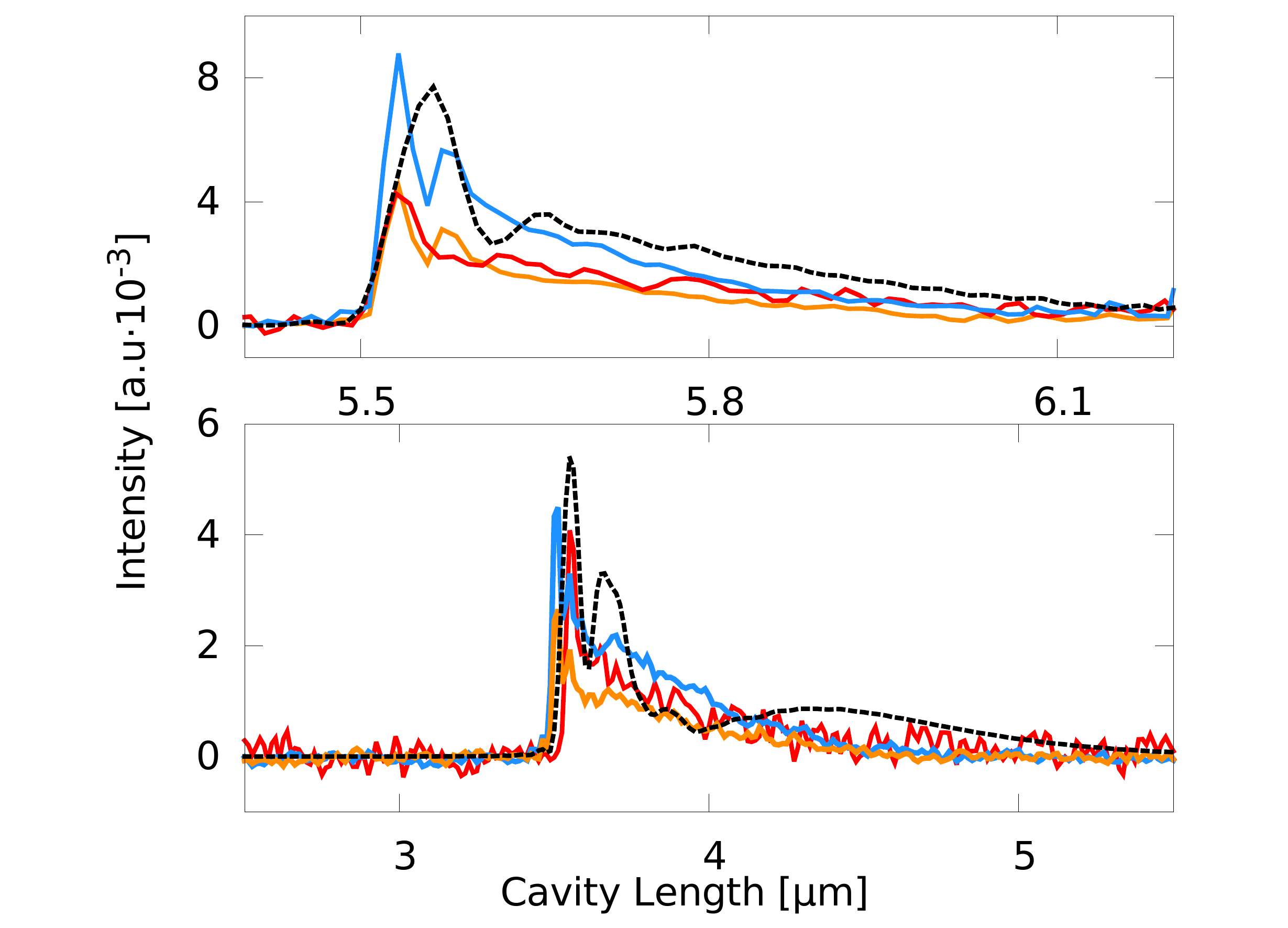}}
\caption{A zoom-in onto the wave fronts of Fig.\ref{AB7} (same color code) for time $t = 100~a.u.$ (upper panel) and $t = 2100~a.u.$ (lower panel). }
\label{AB8}
\end{figure}
\begin{figure}[h!]{
\includegraphics[width=1\linewidth]{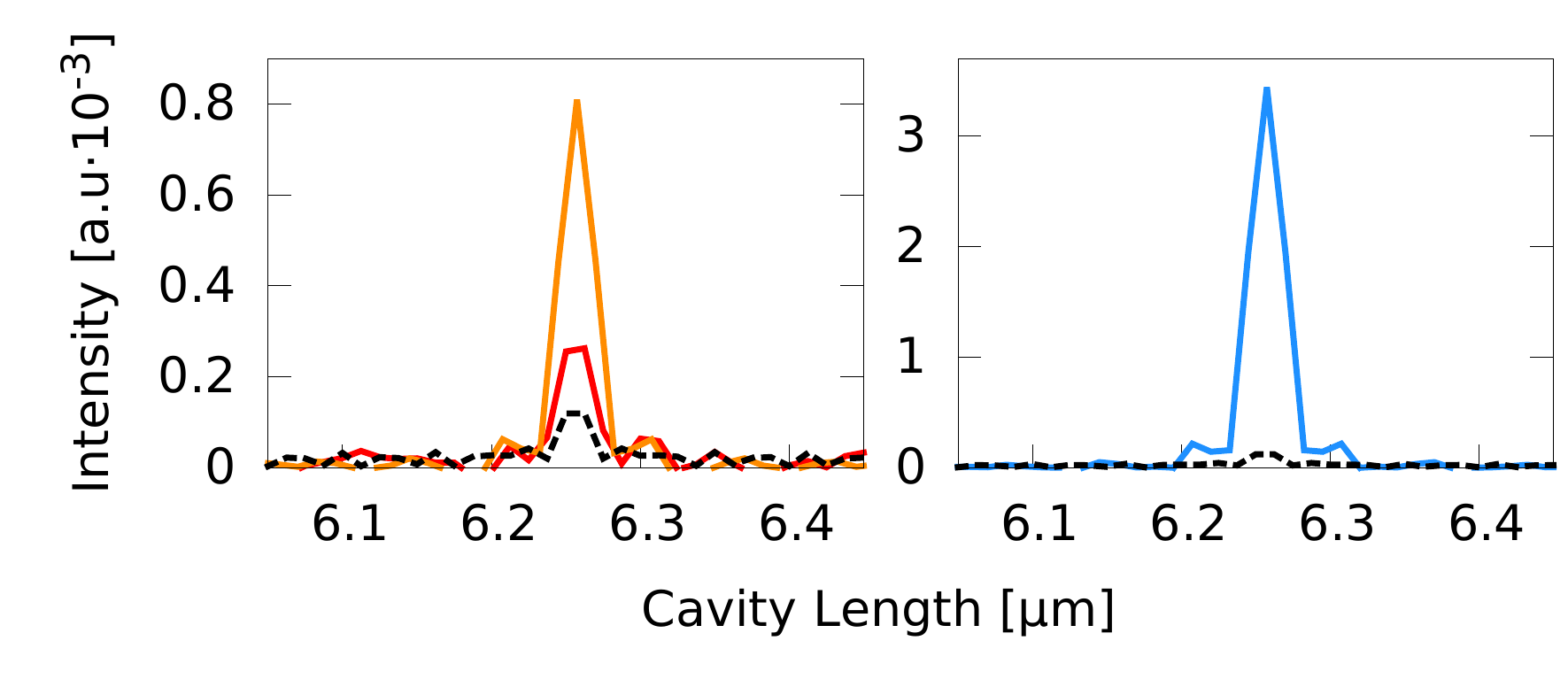}}
\caption{A zoom-in on the bound photon state of Fig.~\ref{AB7} (same color code).}
\label{AB9}
\end{figure}

\begin{figure}[h!]{
\includegraphics[width=1.1\linewidth]{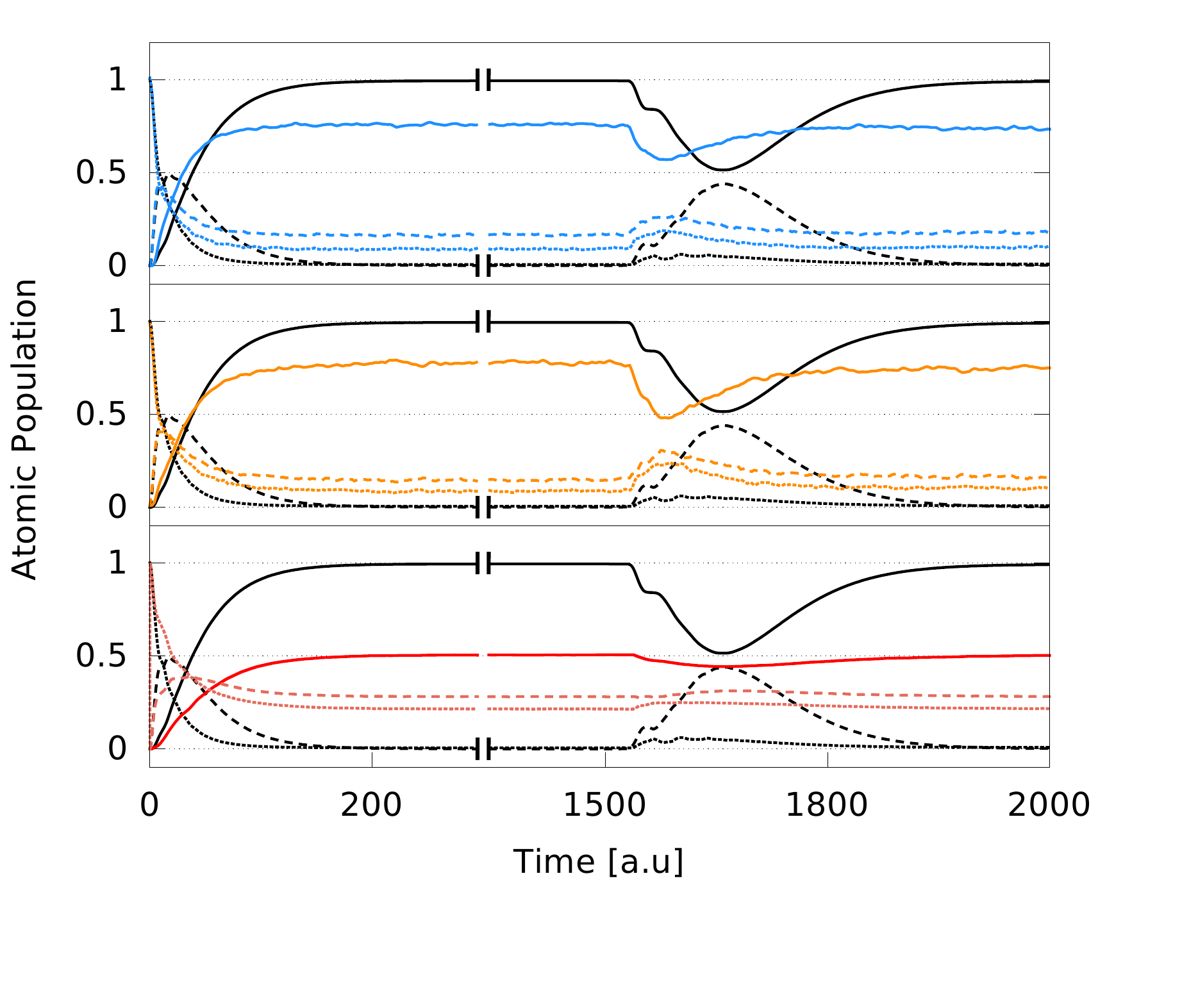}}
\caption{Time-evolution of the atomic state population in the same color-code as Fig.~\ref{AB7}. Solid lines represent the atomic ground state, dashed lines the first excited state and dotted lines the second excited state.}
\label{AB10}
\end{figure}

In Fig.~\ref{AB7} we show the intensity of the cavity field during the two-photon emission process for MTEF, LSC and FBTS compared to the exact solution. Furthermore, in order to allow a more quantitative and accurate comparison, a zoom-in of Fig.~\ref{AB7} is depicted in the same color-code in Figs.~\ref{AB8}, and \ref{AB9} . Here similar dynamics are observed compared to the two-level case. However, due to the additional intermediate atomic state, we now observe a double-peak feature in the emitted photonic wavepacket. This feature corresponds to the emission of two photons, as the excited atom initially decays to the first excited state emitting one photon, and then further relaxes to the ground state, emitting a second photon. 
We find in accordance with the two-level case that the shape of the FBTS-method is in a good agreement with the exact wave-packet shape for time $100$ [a.u.], while the MTEF and LSC-simulation are qualitatively in line, but underestimate the wave-packet amplitude. Further, we observe that at time $2100$ [a.u.] none of the methods sufficiently captures the complex re-emission structure while overestimating the bound photon peak in Fig.~\ref{AB9}.

In Fig.~\ref{AB10} we show the time evolution of the atomic state populations. As before, the emitted photonic wavepacket moves through the cavity, is reflected at the mirrors, and returns to the atom. The first and second excited state are then repopulated due to stimulated absorption. A second spontaneous emission process ensues, and the emitted field again takes on a more complex profile due to interference. While MTEF features the pronounced incomplete emission, LSC and especially FBTS quite accurately capture the short-time decay dynamics. Each method provides a qualitative indication of the re-absorption and consecutive emission with LSC and FBTS performing clearly superior, suffering from a diminished incomplete (de)excitation in relation to MTEF.

\subsection{Computational Effort and Scaling}
Regarding the BBGKY-method the computational cost for this specific model is similar to the exact time-propagation for a two-photon subspace. This makes BBGKY, also in relation to the highly accurate results it provides, the most rigorous method for the model when considering the finite size corrections. Depending on the selected approximation and numerical details such as sparsity, it however features a rather unfavourable high-order polynomial scaling which restricts this method to comparable small systems.

In terms of the other semiclassical approaches, we have found that different numbers of trajectories are needed to converge different observables to the same statistical accuracy. In particular, for subsystem observables like the atomic populations the FSSH and MTEF data are relatively well converged with $10^3$-$10^4$ trajectories, while LSC and FBTS require ~$10^4$ - $10^5$. However, for observables related to the photon field, such as the intensity, the observable remains rather noisy for all the trajectory-based simulation methods with $10^5$ trajectories.

As all the independent trajectory based methods employ a Monte Carlo sampling procedure, their statistical error is proportional to the inverse square-root of the number of trajectories in the ensemble. However, as shown in this work, we have observed that more trajectories are required to converge photon-field (environmental) quantities compared to atomic (subsystem) quantities to within the same relative error. Further, as the trajectories are not coupled during their time evolution, the corresponding algorithms can be implemented in a highly parallel manner to reduce the total run-time.


\section{Conclusion}\label{Se4}

In this work we have adapted and benchmarked a variety of approximate quantum dynamics methods, i.e., multi-trajectory Ehrenfest (MTEF), linearized and partially linearized semiclassical mapping (LSC and FBTS) methods, Tully's fewest switches surface hopping (FSSH), as well as a set of finite size corrected second Born BBGKY truncations, to treat correlated electron-photon systems. We have applied these methods to model QED cavity bound atomic systems in order to simulate the one and two photon spontaneous emission and interference processes, and to analyze the performance of these approaches. 

Consistently for the one and two-photon emission processes, we find that MTEF, LSC and FBTS are able to qualitatively characterize the correct dynamics. The initial spontaneous emission, the associated atomic occupations and emitted photon wavepacket improve from qualitative agreement within MTEF, to slightly better agreement while overestimating the decay-rate in LSC, to almost quantitative agreement using FBTS. However, these methods perform poorly when interference patterns emerge in the reabsorbed and remitted photonic wavepacket; MTEF totally fails to capture any of the interference effects associated with the excitation and re-emission processes, while LSC and FBTS qualitatively recover some of the characteristics of the outgoing intensity. The FSSH-method in contrast is not capable of properly resembling the wavefront of the photonic wavepacket, and furthermore exhibits an incorrect time delay in the re-emitted wavepacket. Consequentially this technique performs rather poorly compared to the other trajectory based methods. It is possible, however, that improved versions of this algorithm may offer improvement over these initial results. The self-consistent perturbative expansion form of the BBGKY-hierarchy behaves exceptionally well when restricted to the physical subspace, although some unphysical effects such as negative photon intensities can result. Finally, all methods investigated here can, in fact, capture the bound photonic state. Here MTEF and BBGKY present the best performance while LSC and FBTS consistently overestimate the amplitude of this feature.

For the two-photon emission process we focused on the most promising approaches considering the balance between performance and computational scalability. Here we find in accordance to the two-level system that  MTEF, LSC and FBTS are able to qualitatively characterize the correct dynamics of this process, however suffer from quantitative drawbacks, especially pronounced for interference features.

Moreover, as experimental advances drive the need for realistic \emph{ab initio} descriptions of light-matter coupled systems, trajectory-based quantum-classical algorithms emerge as promising route towards treating more complex and realistic systems. More precisely, extending to molecular systems beyond the few-level description and incorporating ionic dynamics.
In particular, combining the \textit{ab initio} light-matter coupling methodology recently presented by Jest\"adt et al.\cite{JRORA18} with the multi-trajectory approach could provide a computationally feasible way to simulate photon-field fluctuations and correlations in realistic three-dimensional systems, and work along these lines is already in progress.


\section{Acknowledgements}
We would like to thank J. Flick and N. T. Maitra for insightful discussions and acknowledge financial support from the European Research Council (ERC-2015-AdG-694097). AK acknowledges support from the National Sciences and Engineering Research Council (NSERC) of Canada.

\bibliographystyle{unsrt}
\bibliography{mixed-quantum-methods} 


\end{document}